\documentclass{ws-ijmpa}

\begin{document}

\markboth{T. Sogo, G. R\"opke, P. Schuck}
{Many Body Theory for Quartets, Trions, and Pairs}

%
\catchline{}{}{}{}{}
%

\title{Many Body Theory for Quartets, Trions, and Pairs in Low Density
Multi-Component Fermi-Systems
\footnote{This work is part of an ongoing collaboration 
with Y. Funaki, H. Horiuchi, A. Tohsaki, T. Yamada.}}

\author{P. Schuck}

\address{Institut de Physique Nucl\'eaire, 
 CNRS-IN2P3, UMR8608, F-91406 Orsay Cedex, France\\
schuck@ipno.in2p3.fr}

\address{Universit\'e Paris-Sud, F-91406 Orsay Cedex, France}

\address{Laboratoire de Physique et 
Mod\'elisation des Milieux Condens\'es, 
CNRS and Universit\'e Joseph Fourier, 
Maison des Magistères, Boite Postale 166, 
38042 Grenoble Cedex, France}

\author{T. Sogo, G. R\"opke}

\address{Institut f\"ur Physik, Universit\"at Rostock,
D-18051 Rostock, Germany}

\maketitle


\begin{abstract}
A selfconsistent many body approach for the description of gases 
with quartets, 
trions, and pairs is presented. Applications to 3D Fermi systems at 
low density are discussed.
\end{abstract}

\ccode{PACS numbers: 21.60.Gx, 21.65.-f, 67.85.Lm}

\section{Introduction}

Cluster formation in strongly interacting Fermi-systems is one of the most
interesting subjects in many body physics. So far, in condensed matter, only
the study of formation of two body clusters has been studied in a wide
perspective in the context of pairing, that is superfluidity and
superconductivity (see e.g.\cite{zwerger} in the context of cold atoms). 
However, in nuclear physics, due to the
existence of four different fermions, all attracting one another, the
scenario of cluster formation is much richer. Of course, there also exist
Cooper pairs of neutrons and protons, and to a lesser degree proton-neutron
pairs, but in addition there exist tritons, helions (trions) and
$\alpha$-particles (quartets), to cite only the lightest ones. For instance,
the formation and condensation of $\alpha$-particles is presently in the
focus of our studies. However, adding to a gas of $\alpha$-particles two
neutrons per $\alpha$-particle, at some higher excitation energy, this
system may be transformed into a gas of tritons. For example, one may
imagine that $^{12}$Be is composed out of four tritons at sufficiently high
energies. Also nucleons themselves are trions when seen as  bound states
formed out of three quarks. Then the study of the transition from a
Fermi gas of quarks to a new Fermi gas of trions, i.e. nucleons is interesting.

Recently in cold atoms physics the trapping of three different fermions has
been achieved by two experimental groups and the formation of a gas of 
trions seems possible  \cite{huck}.  May
be in the future, one will even capture four different fermionic atoms and
study quartet condensation. Theoretical work on trions and quartets already has
appeared in quite a number (see articles in  \cite{lechem} and references 
therein) and it seems that the field will rapidly 
expand in the future.  The theoretical works mostly have been 
performed in one dimensional (1D) models. In this work, we will try to 
develop a many body approach for the treatment of a gas of clusters of 
fermions valid in 3D.

\section{Alpha-particle Condensation}

The condensation of $\alpha$-particles can be treated in analogy to the
pairing case. To this end one has to consider the equation for the 
four fermion order parameter $\Psi_{1234}$
 (summation convention is used) \cite{roepke}:
\begin{equation}
\Psi_{1234} = \frac{1-n_1-n_2}{4\mu-\epsilon_{1234}} 
{v}_{121'2'}\Psi_{1'2'34}        + permutations
\label{eq1}
\end{equation}
where $\mu$ is the chemical potential and $\epsilon_{1234} =
\epsilon_1+\epsilon_2+\epsilon_3+\epsilon_4$ where the $ \epsilon_i$'s are the
single particle energies. Besides the matrix elements of the interaction
${ v}_{121'2'}$, the most important ingredients are the single particle
occupation numbers $n_i =  \langle c^+_i c_i \rangle$. 
They, in principle, have to be
calculated from a Dyson equation with a mass operator containing the quartet
condensate. Equation (\ref{eq1}) then constitutes the quartet 
"gap-equation" \cite{roepke}.

A lowest order expression for the single particle mass operator entering the 
Dyson equation can be obtained in the following way where the four particle 
$T$-matrix $T_4$, in the one pole approximation corresponding to (\ref{eq1}), 
is convoluted with three uncorrelated hole lines.
\begin{equation}
\Sigma(1, z_{\nu})= \sum_{234,\Omega_4}T_4(1234,1234;\Omega_4)
G_3^0(234,\Omega_4-z_{\nu})
\label{eq2}
\end{equation}

\noindent
where we used summation over the Matsubara frequency and the uncorrelated 
three body Matsubara Green's function is given by

\begin{equation}
G_3^0(234, z_{\nu}) = \frac{(1-f_2)(1-f_3)(1-f_4) + f_2f_3f_4}
{z_{\nu}-\epsilon_2-\epsilon_3-\epsilon_4}
\label{eq3}
\end{equation}

However, in order to simplify in a first step, we linearize and replace, at
finite temperature, the $n_i$ by the Fermi-Dirac occupation numbers $n_i 
\rightarrow
f_i = [1+e^{(\epsilon_i-\mu)/T}]^{-1}$. Still (\ref{eq1}) is very difficult to
solve, since it is a four body problem rendered more complicated by the
presence of the Pauli blocking factors. However, for clusters involving more
than two fermions, and certainly for strongly bound quartets like the
$\alpha$-particle a projected product ansatz \cite{strsbrg} for 
$\Psi_{1234}$ is known to be a good approximation:
\begin{equation}
\Psi_{1234} \to \delta({\bf K}-{\bf k}_1-{\bf k}_2-{\bf k}_3-{\bf k}_4)
\phi_0({\bf k}_1) \phi_0({\bf k}_2) \phi_0({\bf k}_3) \phi_0({\bf k}_4)
\chi^{ST}
\label{eq4}
\end{equation}
where ${\bf K}$ is the total c.o.m. momentum and $\phi_0({\bf k})$ is a mean
field $0S$-wave function. Spin and isospin are taken care of by $\chi^{ST}$
 with $S=T=0$ for the $\alpha$-particle. We will not consider it further. 
Inserting (\ref{eq4}) into (\ref{eq1}) (with ${\bf K}=0$, for condensation), 
one straightforwardly 
obtains a Hartree-Fock (HF)-type 
of equation for $\phi_0$ which can be solved for various critical temperatures 
$T=T_c$, while $\mu$ is determined from the particle number condition 
$N/V= 4\int \frac{d^3 p}{(2\pi \hbar)^3} f(\epsilon_p)$,
see elsewhere~\cite{spr} for the detailed calculation. For simplicity we 
only take a spin-isospin averaged separable interaction ${ v}_{121'2'}=
\lambda w(\frac{{\bf k}_1-{\bf k}_2}{2}) w(\frac{{\bf k}_1'-{\bf k}_2'}{2})
\delta({\bf k}_1+{\bf k}_2-{\bf k}_1'-{\bf k}_2')$ with the form factor 
$w(k) = e^{-k^2/k_0^2}$.The two open parameters $\lambda$ = -991 MeV fm$^3$ 
and $k_0$ = 1.43 fm$^{-1}$ are 
adjusted that energy (-28 MeV) and radius (1.71 fm) of $\alpha$-particle 
come right. In Fig. 1 
we show the result for $T_c$ as a function of $\mu$ and the density $n$ 
where the free gas Equation of State (EOS) has been used to relate $n$ and 
$\mu$. It is very remarkable that the obtained results for $T_c^{\alpha}$ 
well agree with a direct solution of (1) using a realistic NN force 
\cite{beyer}. 
These results for $T_c^{\alpha}$ are by about 25 percent higher than the ones 
of our earlier publication \cite{roepke}. We, however, checked that the 
underlying 
radius of the $\alpha$-particle in that work is with 2 fm larger than the 
experimental value and that $T_c^{\alpha}$ decreases with increasing radius of 
$\alpha$. Furthermore a different variational wave function was used in 
\cite{roepke}.

In Fig. 1 we also show the critical temperature for deuteron condensation. In 
this case we take $\lambda$ = -1305 MeV fm$^3$ and $k_0$ = 1.46 fm$^{-1}$ to 
get experimental energy (-2.2 MeV) and radius (1.95 fm) of the deuteron. It is 
seen that at higher densities deuteron condensation wins over the one of 
$\alpha$-particles. This stems from the fact that Fermi-Dirac distributions 
in the four body case, see (\ref{eq1}), can never become step-like, as in the 
two body case at zero temperature, since the pairs in an $\alpha$-particle are 
always in motion. As a consequence, $\alpha$-condensation only exists as a BEC 
phase and the weak coupling regime is absent. \\

\begin{figure}[pb]
\centerline{\psfig{file=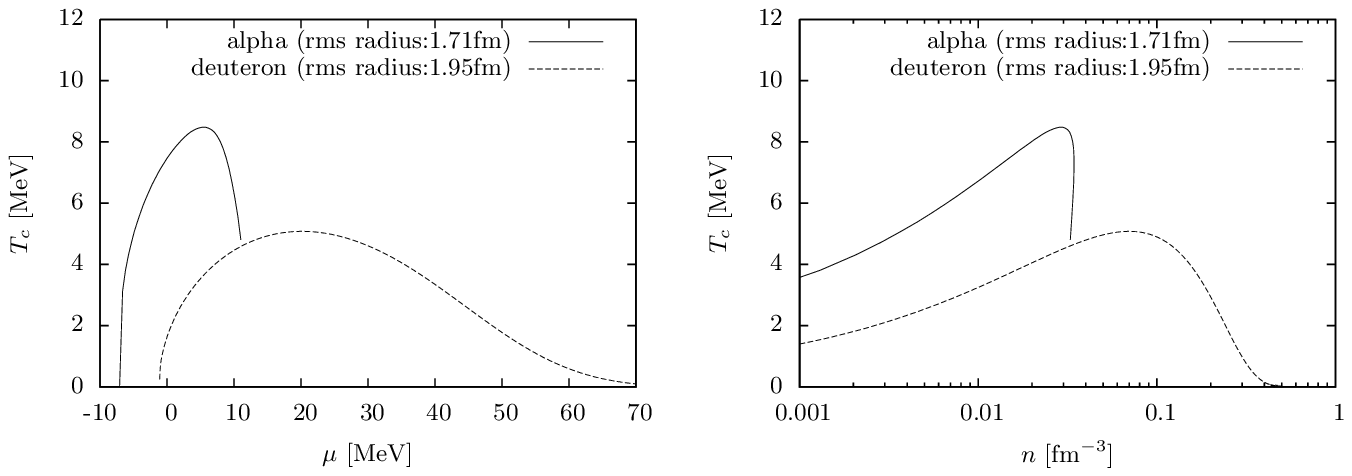,width=10cm}}
\vspace*{8pt}
\caption{Transition temperature of alpha and deuteron
as function of chemical potential (left) and density (right).}
\end{figure}

The mean field ansatz (\ref{eq4}), of course, simplifies the problem 
of quartet 
condensation enormously. In the future it, therefore, should be possible to 
solve the quartet gap-equation (\ref{eq2}) with (\ref{eq4}) fully self-
consistently. At zero temperature this will correspond 
(approximately) to the 
minimization of the energy with a four body coherent state

\begin{equation}
|\Psi_{\alpha}\rangle \sim e^{\frac{1}{4!}\sum_{1234} \phi(1234) 
c_1^+c_2^+c_3^+c_4^+}|vac \rangle
\label{eq5}
\end{equation}

The relation between $\Psi_{1234}$ and $\phi(1234)$ is not trivial but can 
be established in an analogous way to BCS theory.

Condensation phenomena can have precursor signs in finite systems, even small. 
We know that, e.g., from nuclear pairing. 
Evidence for $\alpha$-particle condensation exists in light self conjugate 
nuclei. For example the so-called Hoyle state at 7.65 MeV in $^{12}$C turns 
out to be a product state of three $\alpha$-particles in identical $0S$ orbit 
to over 70 percent \cite{erice}. Indications from theoretical 
results have been put 
forward recently that the 6-th $0^+$ state at 15.1 MeV in $^{16}$O is a four 
$\alpha$-particle condensate state \cite{16O}. If this prediction is further 
verified experimentally and theoretically, then $\alpha$-particle condensation 
very likely is a general phenomenon in nuclear systems \cite{erice}. 
Of course, for 
the description of the condensation of a very small number of $\alpha$-
particles a number conserving variant of (\ref{eq5}) must be employed. 
Also the c.o.m. 
motion is not described by plane waves as in (\ref{eq4}) for 
homogeneous systems but 
the c.o.m. orbitals obey themselves a mean field equation. The corresponding 
ansatz, therefore, is

\begin{equation}
\langle {\bf r}_1, {\bf r}_2, ....|\Phi_{\alpha} \rangle \sim 
{\mathcal A}[\phi(1234)\phi(5678)....]
\label{eq6}
\end{equation}

\noindent
with $\mathcal A$ the antisymmetriser and $\phi(1234)= \Phi_0({\bf R}_{1234}) 
\varphi_0({\bf r}_{ij})$ where ${\bf R}_{1234}=({\bf r}_1+{\bf r}_2+{\bf r}_3+
{\bf r}_4)/4$ is the c.o.m. coordinate and ${\bf r}_{ij} = {\bf r}_i -
{\bf r}_j$ stands for all the possible combinations of relative coordinates of 
the quartet. Of course, at the end the dependence on the total c.o.m. 
coordinate of the system $X_G=[{\bf R}_{1234} + {\bf R}_{5678} + ...]/
n_{\alpha}$, 
with $n_{\alpha}$ the number of quartets has to be eliminated , that is 
one has to 
project on good total c.o.m. momentum zero. In the case of a Gaussian ansatz 
for $\Phi_0({\bf R})$ and $\varphi_0({\bf r}_{ij})$, this is easy to write 
down and we obtain

\begin{equation}
\langle {\bf r}_1, {\bf r}_2, ....|\Phi_{\alpha}\rangle
 \sim {\mathcal A}[\Phi^{c.o.m.}_0 
\phi_0^{intrinsic}]
\label{eq7}
\end{equation}

\noindent
with

\begin{equation}
\Phi^{c.o.m.}_0 
= e^{-({\bf R}_{1234}^G)^2/B^2} e^{-({\bf R}_{5678}^G)^2/B^2}....
\label{eq8}
\end{equation}

\begin{equation}
\phi_0^{intrinsic}= e^{-{\bf s}_{1234}^2/b^2}e^{-{\bf s}_{5678}^2/b^2}....
\label{eq9}
\end{equation}

\noindent
where ${\bf R}_{1234}^G= {\bf R}_{1234} - {\bf X}_G$, and ${\bf s}^2_{1234}=
({\bf r}_1 - {\bf r}_2)^2 + ({\bf r}_1 - {\bf r}_3)^2 + ... .$

We see that the $\alpha$-condensate wave function consists in the anti
summarized product of two interdependent mean field parts, one concerning 
the intrinsic part 
of each $\alpha$-particle and one concerning the c.o.m. motion of the 
$\alpha$-particles. Each part is a mean field wave function with the c.o.m. 
part eliminated. The wave function (\ref{eq6}) is, therefore, 
totally translational 
invariant. This type of wave function is known in the literature as THSR 
wave function \cite{erice}. It depends on two parameters $B$ and $b$ which 
in \cite{erice} were taken 
as Hill-Wheeler coordinates to quantize the corresponding energy surface 
$E_0(B,b)$. For $B \gg b$ one can neglect the antisymmetrizer in (\ref{eq6}) 
and, as  
we see from (\ref{eq6}), the wave function becomes a pure 
product state of $\alpha$-
particles, i.e. a condensate. On the other hand for $B=b$ (\ref{eq6}) is 
equivalent to a pure Slater 
determinant \cite{erice}. The ansatz (\ref{eq6})-(\ref{eq8}) gives 
excellent results for all observables 
of the Hoyle state with the Pauli principle only acting mildly \cite{erice}. 
It also 
predicts similar $\alpha$-condensate states for $^{16}$O, $^{20}$Ne, ... close 
to the corresponding threshold energies for $\alpha$-particle break up 
\cite{16O}.
For the case where in nuclei many $\alpha$-particles existed (e.g. $^{40}$Ca 
with 10 $\alpha$-particles), one also must employ the coherent state 
formulation, since the explicit antisymmetrization becomes very difficult.

\section{Trions}

The case of a gas of trions has not been considered very much in nuclear 
physics.  $^6$He has an excited state at 
$\sim$ 12 MeV which can be interpreted as mainly consisting out of two tritons 
\cite{aoyama}. One can speculate that, e.g. in $^{12}$Be there 
exists a state around 
30 MeV excitation energy with four tritons. Since tritons are fermions, the 
corresponding wave functions of the c.o.m. motion will develope nodes as in 
the usual shell model with $0S, 0P,$ etc. orbitals. The THSR wave function for 
a few tritons may have the form

\begin{equation}
|nt \rangle \sim {\mathcal A}[\phi(123)\phi(456)....]
\label{eq10}
\end{equation}

\noindent
with

\begin{equation}
\phi(123) = e^{-{\bf R}_{123}^2/B^2}\varphi_t({\bf r}_{ij})
\label{eq11}
\end{equation}

\noindent
with definitions analogous to the ones of the quartet case. As already 
mentioned, the antisymmetrizer will make out of the product of c.o.m. 
Gaussians a Slater determinant, whereas the intrinsic wave functions of the 
tritons may stay essentially unaltered.\\

In the nuclear case, a gas of tritons will contain only a very small number of 
such clusters. Considering nucleons as trions formed by three quarks, the 
number of trions can be considerable, that is the number of nucleons in a 
nucleus or in a Heavy Ion Collision. In the case of cold atoms where, 
as mentioned, recently trapping of 
three different fermions has been achieved \cite{huck}, the number of 
trions may go 
in the ten thousand. In such cases the trion wave function $|nt\rangle$ 
of above 
cannot 
be handled directly and we must elaborate some approximate scheme. To put a 
trion creation operator in the exponent like for the quartet is not possible, 
unless one introduces Grassman algebra, since we deal with fermionic clusters. 
One way to 
proceed is via the well known Equation of Motion (EOM) method , in analogy 
to Self Consistent Random Phase Approximation for two 
particles \cite{brueckner}. In this case one first defines in the present 
case a three 
particle creation operator

\begin{equation}
Q_t^+ = \sum_{123}\chi^t_{123}C^+_{123}
\label{eq12}
\end{equation}

\noindent
with $C^+_{123} = c^+_1c^+_2c^+_3 / [(1-n_1)(1-n_2)(1-n_3) +n_1n_2n_3]^{1/2}$ 
the product of three fermionic single 
particle creators. The equation for the amplitudes $\chi$ can be obtained by 
minimizing the following energy weighted sum rule:

\begin{equation}
E_t = \frac{\langle [Q_t,[H,Q_t^+]]_+\rangle}{\langle[Q_t,Q_t^+]_+\rangle}
\label{eq13}
\end{equation}

This leads to a non linear hermitian secular equation for the $\chi$'s. The 
nonlinearity stems from the fact that in the double commutator in 
(\ref{eq13}) appear 
three-body correlation functions of the type $\langle c^+c^+c^+ccc\rangle$. 
The relation 
between $Q_t^+$ and $C^+_{123}$ can be inverted and the three body correlation 
function be expressed by the $\chi$'s with the help of the relation 
$Q_t|0 \rangle = 0$. 
However, two body and one body correlation functions appear as 
well. Though approximate formulas expressing those lower rank correlation 
functions by the $\chi$-amplitudes may be derived, thus obtaining a fully self 
consistent system of equations, the procedure seems extremely heavy. A strong 
simplification could consist in factorizing all correlation functions into an 
antisymmetrized product of single particle correlation functions (single 
particle density matrices) and express 
them via $\chi$-amplitudes in a similar way as we discussed above 
for the alpha 
particle condensation. This procedure is generally known as the so-called 
renormalized RPA, see, e.g., \cite{brueckner}.

On the other hand, if in the gas the trions stay more or less compact, one may 
try a similar mean field ansatz as for the $\alpha$-particles, inspite of the 
fact that a mean field description of only three fermions may not be as good 
as the one for a quartet. However, in the case of three colors, the three 
fermions can occupy the lowest $0S$ level of the mean field and with a 
projection on good total momentum the description may still be quite 
reasonable. 
For instance, 
this may be a quite attractive approach for the constituent quark model where 
the three quarks in a nucleon are bound by a harmonic oscillator field. Let 
us, therefore, make a THSR ansatz \cite{erice} also for the trion wave 
function for the 
case of homogeneous infinite matter

\begin{equation}
\chi^t_{123} \to e^{{\bf K}{\bf R}_{123}}
\delta({\bf K}-{\bf k}_1 -{\bf k}_2 -{\bf k}_3)
\varphi({\bf k}_1)\varphi({\bf k}_2)\varphi({\bf k}_3)\zeta_{123}
\label{eq14}
\end{equation}

\noindent
where $\zeta_{123}$ takes care of spin, isospin, and color indices. 
With (\ref{eq14}) we can 
insert this trial wave function into (\ref{eq13}) and vary with respect to 
$\varphi({\bf k})$. Since the trions are fermions, we also must take care of 
the fact that the trions should carry different c.o.m. momenta ${\bf K}$ 
and, therefore, in (\ref{eq12}) a summation over the c.o.m momenta must be 
contained. 
The case of trion formation has been studied theoretically in solving 
exactly 1D-model Hamiltonians with three colors \cite{lechem}. 
It would be interesting 
to compare the results of above approach to these exact solutions.

In the case of nucleons, one pair of quarks is forming a strongly correlated 
di-quark state. It seems reasonable to approximate the di-quark by a boson. 
Then the transition of a gas of quarks and di-quarks can be considered as a 
mixture of fermions and bosons. The nucleons then constitute a bound state 
between a fermion and a boson. Interesting physics has recently been found in 
treating the fermion-boson scattering in the background of a gas of bosons and 
fermions (the one pair approximation). Most interestingly the fermion-boson 
pair seems to form a stable pair even for infinitesimal attraction, quite 
similar to what happens for the formation of Cooper pairs in a two component 
Fermi gas. However, the boson-fermion pairs have, of course, Fermi statistics, 
thus forming a new Fermi gas of composites \cite{bf}.

\section{Pairs}

In the pairing case, the focus is on the BEC-BCS transition. 
Researches of BEC-BCS crossover have been done 
not only for the condensed matter~\cite{becbcs} but also 
for the nuclear matter~\cite{roepke2}.
The application 
of the Nozi\`eres Schmitt-Rink approach \cite{nsr} 
is fairly standard by now, taking 
care of the purely bosonic (strong coupling) limit of the pairs and 
their c.o.m. motion in the 
Bose-Einstein distribution. 
However, in the strong coupling limit, the pairs 
behave as (composite) particles and create their own mean field. This effect 
is not taken into account in the work of \cite{nsr}, as 
discussed by the authors 
themselves. In order to include this effect, we can derive two coupled mean 
field equations, one fermionic, the other bosonic (fermion pairs). This is 
achieved by the minimization of the following generalized sum rules

\begin{equation}
\varepsilon_k = \langle [q_k[H,q_k^+]]_+\rangle/\langle[q_k,q_k^+]_+\rangle
\label{eq15}
\end{equation}

\noindent
with $q_k^+=u_kc_k^+ - v_kc_{\bar k}$ being the standard Bogoliubov 
transformation of fermions. The corresponding bosonic equation is obtained from

\begin{equation}
E_{\bf K,\nu} = \langle [Q_{{\bf K},\nu},[H,Q^+_{{\bf K},\nu}]] \rangle/
\langle [Q_{{\bf K},\nu}, Q^+_{{\bf K},\nu}]\rangle
\label{eq16}
\end{equation}

\noindent
where

\begin{equation}
Q^+_{{\bf K},\nu}= \sum_{{\bf k}_1 {\bf k}_2} [X^{{\bf K},\nu}_{{\bf k}_1 
{\bf k}_2} q^+_{{\bf k}_1} q^+_{{\bf k}_2} - 
Y^{{\bf K},\nu}_{{\bf k}_1 {\bf k}_2} q_{{\bf k}_1} q_{{\bf k}_2}]/(1-
n_{{\bf k}_1} - n_{{\bf k}_2})^{1/2} 
\label{eq17}
\end{equation}

\noindent
represents a Bogoliubov transformation for fermion pairs.

\noindent
A momentum conserving delta function between ${\bf K}$ and 
${\bf k}_1, {\bf k}_2$ is implicit and spin and isospin indices have been 
suppressed. The expectation values are with the correlated ground state 
defined (approximately) by $Q|0\rangle=0$.

Equations (\ref{eq15})-(\ref{eq17}) can be generalized to finite temperature 
in introducing corresponding Gorkov equations. The fully selfconsistent set of 
equations for amplitudes $u,v,X,Y$ can be worked out (see \cite{picket} and 
references therein) and their 
solutions give very promising results in non-trivial model cases \cite{picket}.
They constitute a fully selfconsistent extension of the 
Nozi\`eres Schmitt-Rink approach \cite{nsr}.

\section{Conclusions}

Selfconsistent many body approaches for gas phases of quartets, trions, and 
pairs in multicomponent Fermi-systems have been established. They are 
applicable to realistic 3D systems accounting for strong cluster phenomena.

\section*{Acknowledgments}

This work is supported by the DFG grant No. RO905/29-1.
Useful discussions with P. Lecheminant are greatfully acknowledged.


\end{document}